# Beyond the Tumor: Recurrence-Prone Radiomics for Prognostication in Negative PSMA PET/CT scans of Prostate Cancer

Fereshteh Yousefirizi[1*], Sara Harsini[2], Mobin Mohebi[3], Ian Alberts[2,4], Tahir Yusufaly[5], Monica Luo[1,4], Hamid Abdollahi[1,4], Elmira Yazdani[6], Soheila Mirabedin[7], Maziar Sabouri[1,4], Parham Geramifar[8], Peyman Sheikhzadeh[7], Patrick Martineau[2,4], Don Wilson[2], François Bénard[2,4], Carlos Uribe[1,2,4], Arman Rahmim[1,4]

[1] Department of Basic and Translation Research, BC Cancer Research, Vancouver, BC, Canada
[2] Molecular Imaging and Therapy, BC Cancer, Vancouver, BC, Canada
[3] Institut de Biologie Valrose, Université Côte d'Azur, CNRS, Inserm, Nice, France
[4] University of British Columbia, Vancouver, BC, Canada
[5] Russell H. Morgan Department of Radiology and Radiological Sciences, Johns Hopkins School of Medicine, Baltimore, Maryland, USA
[6] Medical Physics Department, School of Medicine, Iran University of Medical Sciences, Tehran, Iran
[7] Department of Nuclear Medicine, IKHC, Tehran University of Medical Science, Tehran, Iran
[8] Research Center for Nuclear Medicine, Tehran University of Medical Sciences, Tehran, Iran

**Abstract**

**Purpose:** Prostate-specific membrane antigen (PSMA) PET/CT is routinely used to restage prostate cancer (PCa) with biochemical recurrence (BCR), yet negative scans may conceal subclinical disease. This study evaluated whether radiomics features from recurrence-prone organs on negative [$^{18}$F]DCFPyL PET/CT can predict clinical progression (CP) and progression-free survival (CPFS).

**Materials and Methods:** A post-hoc analysis was conducted in 132 BCR patients (mean age 74.5 years) with negative [$^{18}$F]DCFPyL PET/CT scans who received no further treatment. Organs commonly involved in recurrence (e.g., prostate bed, lymph nodes, bones, liver, lungs) were segmented using nnU-Net (TotalSegmentator), and radiomics features were extracted. Machine learning (XGBoost) and Cox models were evaluated using nested cross-validation (5-fold outer, 3-fold inner). External validation across two independent centers employed ComBat harmonization. The effect of reader-assigned diagnostic certainty on model performance was also analyzed. We focused only on recurrence-prone organs because prior studies show >80% of PCa recurrences occur in these locations.

**Results:** Median PSA at imaging was 0.74 ng/mL. Over a median 25.5-month follow-up, 42 patients (31.8%) progressed. The C-index improved from 0.65 for clinical variables alone to 0.74 when combining PET, CT, and clinical features (p < 0.05). Diagnostic certainty about negative PSMA PET/CT strongly influenced predictive accuracy with higher performance in moderate-certainty negative scans than in high-certainty scans. Radiomics from negative PSMA PET/CT reflected systemic recurrence risk and localized early lung metastases. External validation showed <10 % performance reduction despite inter-center differences.

**Conclusion:** Radiomics signatures from recurrence-prone organs on negative PSMA PET/CT predict CP and CPFS in BCR patients. These features may capture subclinical alterations undetectable visually and provide complementary biomarkers for risk stratification.

## INTRODUCTION

Prostate cancer (PCa) remains a major global health concern, ranking second in incidence and fifth in mortality among men (1). Despite therapeutic advances, 20–50% of patients develop biochemical recurrence (BCR) (2), managed through salvage radiotherapy (sRT), androgen deprivation therapy (ADT), or surveillance (3-5). Prostate-specific membrane antigen (PSMA) PET/CT is the current gold standard for staging and restaging owing to its high sensitivity (6-11), though lesion detection varies with voxel size, radiotracer, and scanner type (12-14). Identifying false-negative PSMA PET/CT cases remains crucial.

Clinical progression may occur despite negative scans, owing to bladder activity or micrometastases. Although negative PSMA PET/CT often indicates favorable prognosis, it may overlook subclinical disease (15). Current guidelines recommend sRT for high-risk BCR after negative scans (16), yet evidence remains limited, and the optimal use of sRT ± ADT is unclear. Improved risk stratification is needed to distinguish patients suited for surveillance, local salvage, or those likely to develop distant metastases.

Beyond lesion detection, recent studies show that imaging-derived host features, such as muscle and adipose tissue metrics, carry prognostic value in recurrent PCa (17, 18). These findings support exploring quantitative, segmentation-based imaging biomarkers in PSMA-negative cohorts to refine recurrence risk assessment.

Radiomics has enhanced prognostic modeling across cancers (19-22), yet most studies focus on tumor-bearing regions, neglecting the physiological information within normal organs routinely captured in whole-body imaging. Emerging evidence indicates that radiomic features from non-tumor tissues such as fat and bone may reflect systemic disease effects (23, 24). uilding on this concept, we examined whether radiomics features from recurrence-prone organs on negative [$^{18}$F]DCFPyL PET/CT scans could predict clinical progression (CP) in biochemically recurrent (BCR) prostate cancer patients managed by observation. Given that PCa commonly recurs in lymph nodes, bone, and distant organs (25), we hypothesized that radiomics from common sites of recurrence could reveal early systemic changes missed by visual interpretation.

Radiomics features were extracted from organs without visible tumor involvement in patients with biochemical recurrence and negative PSMA PET/CT scans. Our study is the first to show that radiomics from common sites of recurrence on negative PSMA PET/CT scans independently associated with progression risk. This study evaluates whether radiomics from recurrence-prone organs can predict progression risk in PSMA-negative patients. The results are intended to improve biological understanding and risk stratification, rather than to directly guide management decisions.

**METHODS**

Figure 1 illustrates the structured framework employed in our study, comprising four main components: A) study cohort selection, B) image data preprocessing, C) radiomics feature curation, and D) predictive modeling.

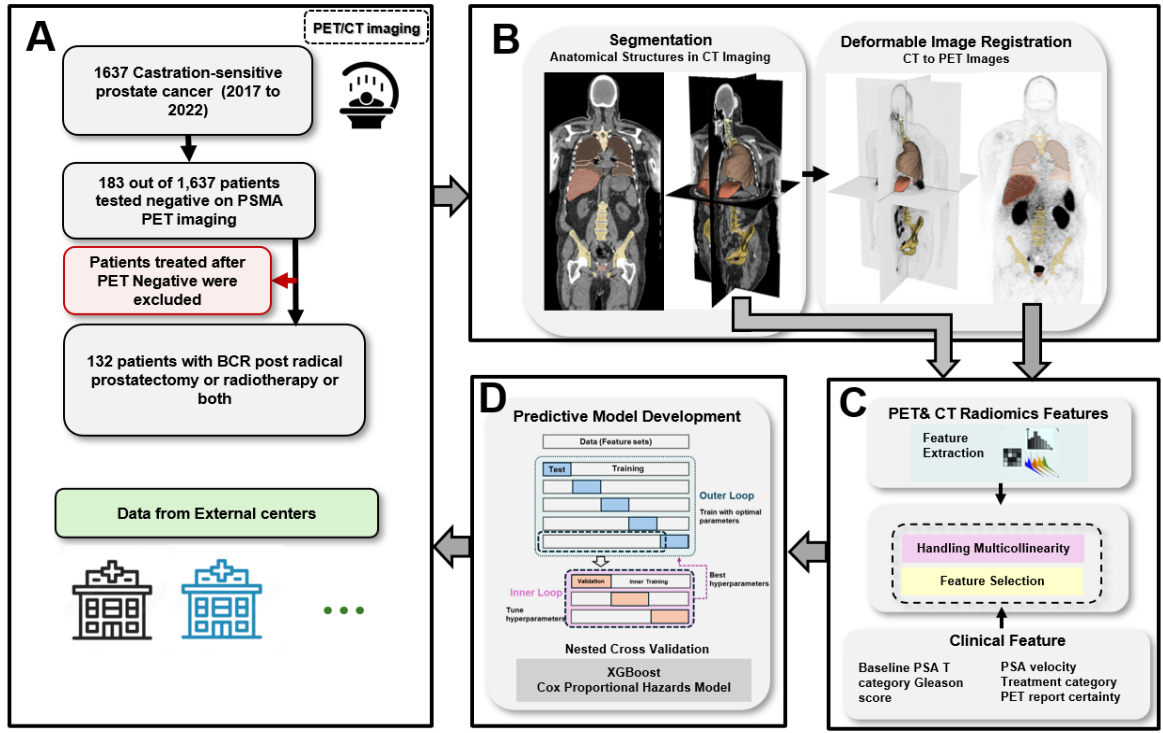

Figure 1: The block diagram of the systematic approach pursued in this study. A: study cohort selection, B: Image Data preprocessing, C: radiomics feature curation, D: Predictive modeling

*Data description*

Center 1: This post-hoc subgroup analysis was derived from a prospective, non-randomized clinical trial (NCT02899312) enrolling men with advanced or recurrent prostate cancer who met one of four criteria: (i) biochemical recurrence (BCR) after prostatectomy (PSA > 0.4 ng/mL), (ii) BCR after radiotherapy (PSA > 2 ng/mL above nadir), (iii) castration-resistant disease (PSA ≥ 2 ng/mL, testosterone < 1.7 nm/L), or (iv) inconclusive imaging. Exclusion criteria included medical instability, inability to consent or lie supine, scanner size limitations, or ECOG > 2. The study was approved by the institutional Research Ethics Board, and all participants provided written informed consent.

For this analysis, only PSMA-negative PET/CT scans (visually and SUV-based) were included. Each participant received weight-adjusted [$^{18}$F]DCFPyL (237–474 MBq) after a four-hour fast. Imaging commenced with 120 min post-injection using GE Discovery 600/690 scanners (GE Healthcare, USA). A low-dose non-contrast CT was acquired for localization and attenuation correction, followed by whole-body PET reconstructed with OSEM and point-spread-function modeling.

Patients from two external centers (2 and 3) with BCR and no further treatment were included. Center 2 contributed 6 [$^{68}$Ga]PSMA PET/CT cases (GE Discovery IQ, 3.26 mm³ voxels; median PSA 1.2 ng/mL) and Center 3 contributed 21 cases (Siemens Biograph mCT, 5 mm³ voxels; median PSA 1.6 ng/mL). Median follow-up was 15.2 and 15.6 months, respectively. Full acquisition and cohort details are provided in the Supplementary File.

*Image Interpretation, Follow-up and Outcome Measures*

No additional treatment was administered after the negative scan. Patients were followed for clinical progression (CP), defined as the radiologic emergence of new lesions on follow-up imaging (PET/CT, MRI, CT, or bone scan). PSA-only progression was not classified as CP. Clinical progression-free survival (CPFS) was measured from the date of the negative scan to radiographic recurrence, with non-progressors censored at last imaging. A three-level certainty score (high, moderate, low) was assigned to each scan by three expert physicians to reflect interpretive confidence and mitigate inter-reader variability.

*Segmentation and Registration*

Organ segmentation was performed using TotalSegmentator (26) built on nnU-Net (27), validated across multiple clinical scenarios (26, 28). PET/CT co-registration was executed using Elastix with B-spline transforms and mutual information similarity metrics (29).

*Feature Extraction*

Prior to radiomic feature extraction, all PET and CT images were resampled to isotropic voxel spacing using the internal resampling function of LIFEx (v25.07.n) to ensure spatial consistency across scans. PET voxel intensities were discretized into 64 fixed bins spanning 0 to SUVmax using absolute resampling, following IBSI guidelines. As analysis was performed on non-tumor, recurrence-prone organ ROIs in PET-negative scans, absolute discretization (0–SUVmax) was used to preserve physiologic uptake patterns while standardizing gray-level sampling across subjects. CT data were clipped to tissue-specific HU ranges (lung –900–0, soft tissue –300–300, bone 0–800) and resampled to 400 gray levels using 10-HU bins. Feature extraction followed the IBSI-compliant (30) LIFEx Texture protocol (31).

Extracted features included intensity, shape, and texture matrices: gray level co-occurrence matrix (GLCM), gray level run length matrix (GLRLM), gray level size zone matrix (GLSZM), gray level dependence matrix (GLDM), neighborhood gray tone difference matrix (NGTDM), and cumulative SUV histogram (CSH) area under the curve. Clinical variables (baseline PSA, staging, Gleason score, PSA kinetics, treatment category) and physician-assigned interpretive certainty (high/moderate/low) (three expert readers) were also integrated into the feature set. The list of the features and related explanations are provided in the supplementary file (Table S1).

### *Feature selection and Machine learning*

Following EANM/SNMMI guidelines for radiomics studies (32), we developed a pipeline integrating clinical and imaging features, evaluated through a nested cross-validation (CV) (5-fold outer, 3-fold inner) to minimize overfitting (33). Feature preprocessing included removal of highly correlated variables (Pearson r > 0.95), standardization to zero mean/unit variance, and two-step feature selection: Lasso regression (L1 penalty) followed by sequential forward selection to retain up to 15 features. Disease progression was modeled using XGBoost classifiers (Python 3.9, XGBoost 1.7.4) (34, 35)), and time-to-event prediction was performed using Cox proportional hazards models with Kaplan-Meier and concordance index (C-index) evaluation. Variance Inflation Factor was used to assess multicollinearity.

Seven feature sets, Clinical, PET, CT, PET+CT, PET+Clinical, CT+Clinical, and PET+CT+Clinical, were tested for predicting disease progression and CPFS. Pairwise Mann-Whitney U tests were used to compare each set against the Clinical baseline, with Bonferroni correction applied (corrected $p < 0.05$). To assess model robustness across heterogeneous clinical contexts, subgroup analyses were conducted based on treatment modality (RP, RT, RP+RT), with independent feature selection and model training in each group. Additionally, reader-assigned interpretive certainty categories (high/moderate/low) were incorporated to account for interpretive variability in PET-negative scans.

### *Organ-Specific PET Radiomics for Predicting Site-Specific Metastases*

To assess the ability of PET radiomics to predict site-specific metastases, binary classification models were developed using features extracted from anatomically relevant regions. Class imbalance was addressed using SMOTE, and model performance was evaluated via 5-fold stratified cross-validation. Multiple classifiers, including Random Forest and Logistic Regression, were compared. Feature importance was quantified using permutation-based analysis to identify the most predictive features for each metastatic site.

### *Reader certainty Analysis*

To evaluate the influence of interpretive confidence on model performance, each PET/CT scan was retrospectively assigned a three-level certainty score (high, moderate, low) by three independent nuclear medicine physicians (with the average 12 years of experience). A 5-fold outer cross-validation framework was employed, incorporating LASSO-based feature selection and sequential forward selection within each fold. Standardized features were used to train XGBoost classifiers for clinical progression prediction, and Cox proportional hazards models were fitted for CPFS estimation. Performance metrics, mean accuracy, ROC-AUC, and C-index, were aggregated across folds to ensure robustness.

*Multicenter External validation*

A standardized pipeline for PET/CT registration, organ segmentation, and radiomics extraction was applied uniformly across all centers. To mitigate scanner-related variability, ComBat harmonization was performed separately for PET and CT features using imaging center as the batch covariate. Harmonization was conducted in an unsupervised manner, excluding outcome labels to reduce bias. Due to variability in clinical data availability across sites, external validation included only imaging-derived radiomics features.

## RESULTS

*Patients and Clinical Data*

Among 1,637 castration-sensitive PCa patients enrolled between March 2017 and September 2022, 183 had negative PSMA PET imaging. After excluding those who received post-PET treatment (sRT or ADT), 132 patients (mean age 74.5 years, range 55–92) with biochemical recurrence after curative-intent therapy (RP, RT, or both) and no further treatment were included. The median PSA at PET/CT was 0.74 ng/mL (range: 0.4–19.4), and the median CPFS was 25.5 months (range: 2–71). During follow-up, 42 patients (31.8%) experienced progression. Recurrence sites included pelvic lymph nodes (42.9%), bone (35.7%), prostatic fossa (30.9%), distant lymph nodes (26.2%), and lungs (14.3%). Baseline characteristics are summarized in Table 1.

Table 1. Surveillance Subgroup Patient and treatment characteristics.

| Characteristic | Entire Cohort |
|---|---|
| Number of patients | 132 |
| Age at PET/CT (years), median (range) | 74.5 (55-92) |
| ISUP grade group, number (percentage) | |
| 1 | 19 (14.3%) |
| 2 | 41 (30.8%) |
| 3 | 34 (25.6%) |
| 4 | 13 (9.8%) |
| 5 | 21 (15.8%) |
| Missing | 4 (3.0%) |
| Primary tumor classification, number (percentage) | |
| p/cT1 | 8 (6.1%) |
| p/cT2 | 59 (44.7%) |
|     p/cT3 | 57 (43.2%) |
|     Missing | 8 (6.1%) |
| Primary nodal status, number (percentage) | |
|     p/cN0 | 89 (67.4%) |
|     p/cN1 | 12 (9.1%) |
|     p/cNx | 25 (18.9%) |
|     Missing | 6 (4.5%) |
| Time from diagnosis to recurrence (months), median (range) | 25.5 (3-71) |
| PSA at PET/CT (ng/ml), median (range) | 0.74 (0.4-19.4) |
| PSA velocity at PET/CT (ng/ml/year), median (range) | 0.5 (-106.4-25.6) |
| Missing | 4 cases (3.0 %) |
| Follow-up time (month) since PET/CT, median (range) | 37 (12-79) |
| Treatment Category | |
|   Radical prostatectomy | 61 (46.2%) |
|   Radiotherapy | 28 (21.2%) |
|   Radical prostatectomy and Radiotherapy | 43 (32.6%) |
| Pre-PET ADT | 43 (32.6%) |
| Physician Certainty in Negative PET report | |
|   High | 74 (56.1%) |
|   Moderate | 27 (20.5%) |
|   Low | 7 (5.3%) |
|   Missing | 24 (18.2%) |

PET/CT, positron emission tomography/computed tomography; ISUP, International Society of Urological Pathology; RP, radical prostatectomy; RT, radiation therapy; ADT, androgen deprivation therapy; PSA, prostate-specific antigen; PSA-DT, prostate-specific antigen doubling time.

## *Modelling and Feature Set Evaluation*

Seven feature sets were evaluated for predicting disease progression and CPFS. The Clinical-only model (e.g., baseline PSA, T category, Gleason score, ISUP grade, PSA velocity, pre-PET ADT, treatment category) yielded the lowest C-index ($0.65 \pm 0.05$). PET radiomics alone improved performance (C-index: $0.72 \pm 0.10$), with key features including lung GLCM Correlation and vertebrae GLSZM Small Zone Emphasis. Combined PET+Clinical yielded a C-index of $0.71 \pm 0.15$. CT radiomics alone achieved $0.66 \pm 0.17$, improving to $0.71 \pm 0.14$ when integrated with clinical data. The highest performance was observed for PET+CT+Clinical (C-index: $0.74 \pm 0.13$), demonstrating the added value of multimodal integration. Pairwise Mann-Whitney U tests (Bonferroni-corrected) showed that PET, PET+Clinical, CT+Clinical, and PET+CT+Clinical significantly outperformed the Clinical model (corrected p < 0.05) (Table S2). Kaplan-Meier curves revealed significant CPFS stratification across risk groups for all feature combinations. (Figure S1)

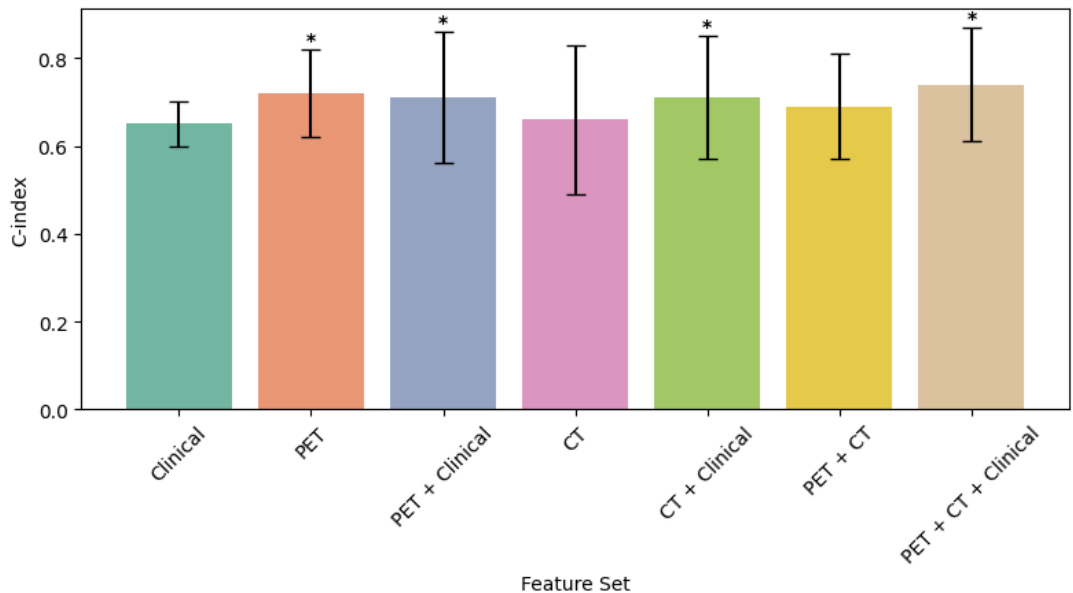

Figure 2. Pairwise comparison of C-index values between Clinical and other feature sets. Statistical significance was determined using Mann-Whitney U tests with Bonferroni correction. PET, PET + Clinical, CT + Clinical, and PET + CT + Clinical showed statistically significant improvements over Clinical alone (corrected p-value <0.05), while CT and PET + CT did not. PET: PET radiomics, CT: CT radiomics

### *Subgroup Analysis by Treatment Category*

Model performance was consistent across treatment subgroups—radical prostatectomy (RP), radiotherapy (RT), and RP+RT, with C-indices ranging from 0.53 to 0.68. No statistically significant differences were observed (Mann-Whitney U test, p > 0.0167), suggesting generalizability of radiomics signatures across treatment contexts while highlighting the importance of subgroup-specific validation (Figure 3).

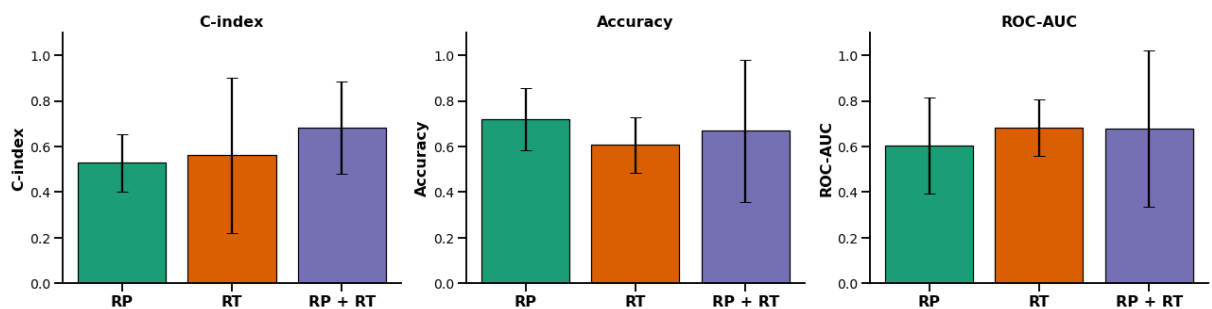

Figure 3. Comparison of model performance across treatment subgroups: RP, RT, and RP + RT. Boxplots show the distribution of C-index, Accuracy, and ROC-AUC across five outer folds for each group. Bars represent mean

values with standard deviations. Models were trained using nested cross-validation with Lasso-based feature selection and XGBoost classification. Time-to-progression was evaluated using Cox proportional hazards modeling. No statistically significant differences were observed between groups (Mann-Whitney U test with Bonferroni correction, $p > 0.0167$).

### Site-Specific Metastasis Prediction with Organ-Specific Radiomics

For metastasis site prediction, Random Forest performed best for bone metastases (ROC-AUC: $0.57 \pm 0.12$), while Logistic Regression yielded the highest performance for lung metastases (ROC-AUC: $0.63 \pm 0.18$). Feature importance analysis revealed that sacral and vertebral features were dominant for bone prediction, while SUV and texture features from lung regions were most informative for pulmonary metastases. While modest, these results suggest potential utility for site-directed radiomics and motivate further validation (Figure 4 (a) and (b)).

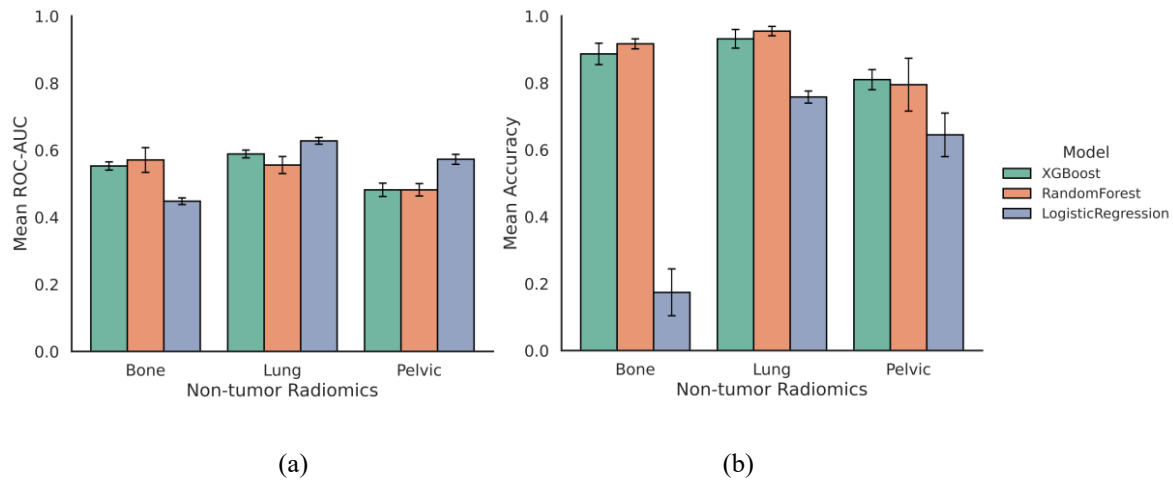

(a)  (b)

Figure 4. Performance comparison of classification models (XGBoost, Random Forest, Logistic Regression) for predicting site-specific metastases using organ-specific PET radiomics features. (a) Mean ROC-AUC per model across bone, lung, and pelvic regions. (b) Mean accuracy per model across the same anatomical regions. Logistic Regression showed the highest ROC-AUC for lung metastasis prediction, while Random Forest achieved the highest ROC-AUC for bone metastases. Accuracy values were generally highest for XGBoost and Random Forest across all regions. These results highlight the anatomical variability in predictive performance and suggest the potential value of region-specific modeling strategies. Error bars indicate SD across outer cross-validation folds.

### Certainty analysis

A subset of 101 patients was stratified by reader-assigned interpretive certainty (74 high, 27 moderate). Certainty was available for 108/132; the low-certainty group (n=7) was excluded, leaving 101 (74 high, 27 moderate). Model performance was significantly higher in moderate-certainty scans than in high-certainty scans (C-index = $0.65 \pm 0.13$ vs $0.35 \pm 0.13$; ROC-AUC

= 0.59 ± 0.15 vs 0.38 ± 0.13; accuracy = 0.53 ± 0.12 vs 0.36 ± 0.12; Bonferroni-corrected p = 0.0358).

These findings indicate that radiomics features from non-tumor regions are most predictive in cases where PET/CT interpretation is uncertain despite negative visual findings. Moderate-certainty scans likely contain subtle physiologic or systemic alterations related to early or subvisual disease that are not confidently identified by expert readers but are captured by radiomic patterns. In contrast, unequivocally negative scans may represent biologically stable states with less informative variation, reducing predictive signal. These results highlight the complementary role of non-tumor radiomics in supporting risk stratification specifically in visually uncertain PSMA-negative cases (Figure 5).

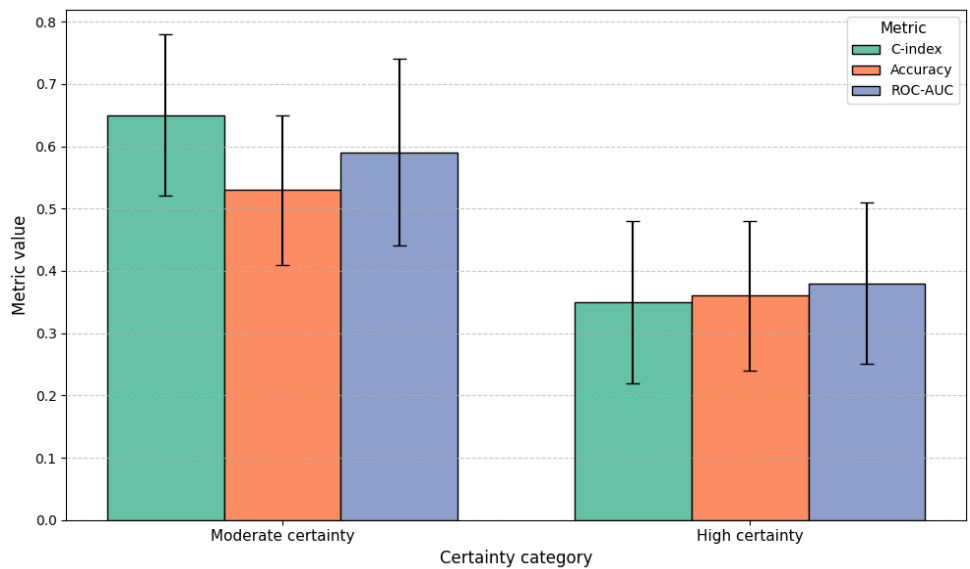

Figure 5. Model performance stratified by reader-assigned interpretive certainty. Radiomics models showed higher performance in moderate-certainty negative scans than in high-certainty scans across C-index, accuracy, and ROC-AUC. Error bars represent standard deviations across cross-validation folds; statistical significance was confirmed after Bonferroni correction (p = 0.0358).

### *External validation*

Despite ComBat harmonization, a performance drop was observed across external centers due to domain shift arising from radiotracer variation ([$^{18}$F]DCFPyL vs. [$^{68}$Ga]PSMA-11), scanner type, and protocol differences. PET+CT C-index: dropped by 10.1% (Center 2) and 7.2% (Center 3). PET-only C-index: dropped by 6.9% and 9.7%, respectively. CT-only C-index: decreased by 9.1% and 1.5%, respectively. These results highlight persistent generalization challenges and the need for harmonized acquisition protocols or domain-adaptive modeling (Table 2).

Table 2. The result of external validation (CV: Cross validation)

| Feature Set | CV C-index (mean ± SD) Center1 | External C-index (Center 2) | Drop % (Center 2) | External C-index (Center 3) | Drop % (Center 3) |
|---|---|---|---|---|---|
| PET+CT | 0.69 ± 0.12 | 0.62 | 10.1% | 0.64 | 7.2% |
| PET | 0.72 ± 0.10 | 0.67 | 6.9% | 0.65 | 9.7% |
| CT | 0.66 ± 0.17 | 0.6 | 9.1% | 0.65 | 1.5% |

Representative patient examples illustrating concordance between model predictions and longitudinal outcomes are shown in Figure 6.

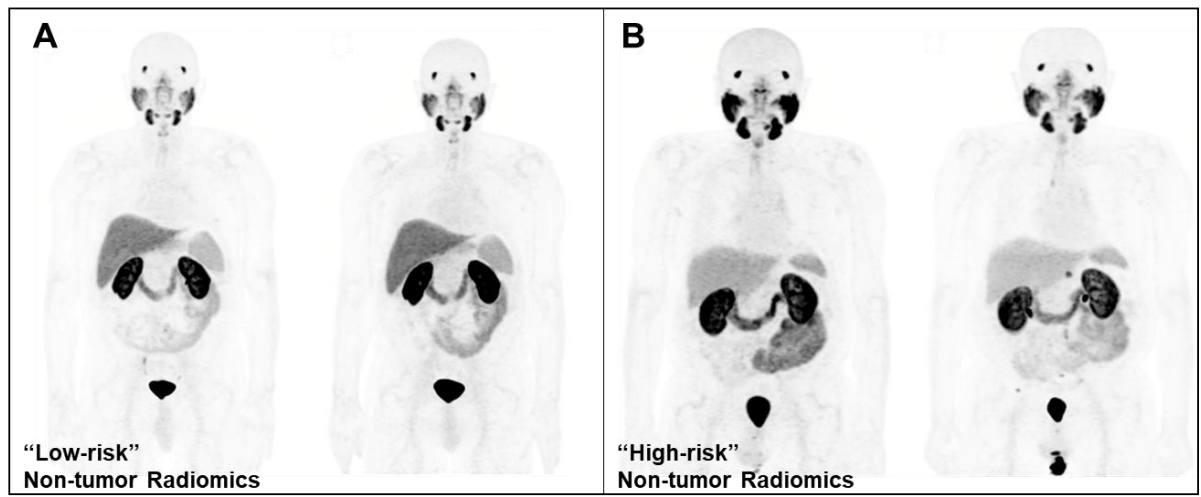

Figure 6. (**A**) A patient with biochemical recurrence after prostatectomy had an initial negative [$^{18}$F]DCFPyL PET/CT. Application of our trained non-tumor PET radiomics model predicted low recurrence risk, and a follow-up scan 25 months later remained negative, confirming agreement between model prediction and clinical outcome. (**B**) A patient with biochemical recurrence had an initial negative [$^{18}$F]DCFPyL PET/CT. Our trained model predicted high recurrence risk, and follow-up imaging 28 months later revealed multiple PSMA-avid lesions, confirming the model prediction and the patient's true outcome.

## DISCUSSION

Early detection of PCa recurrence remains crucial for optimizing treatment, particularly in early BCR where imaging sensitivity is limited. Although PSMA PET/CT markedly improves lesion detection and guides management (36-39), even advanced scanners (14), may miss subclinical disease at low PSA levels. Negative PSMA PET/CT scans are generally linked to favorable outcomes, yet a notable proportion of patients still progress (15, 40-42), underscoring ongoing debate over early salvage radiotherapy versus observation (43-45). Clinical tools such as PSA kinetics and EAU risk stratification provide partial guidance (46), but lack individualized precision. In our cohort, 31.8 % of patients with negative scans experienced progression, often in pelvic nodes or the prostatic fossa. These findings suggest that radiomics from recurrence-prone sites may capture subtle systemic alterations preceding overt metastasis, offering a potential biomarker for global recurrence risk.

Radiomics from common sites of recurrence have shown prognostic value in other cancers, including low-uptake tumors (23, 24, 47). We demonstrate that recurrence-prone radiomics on negative [$^{18}$F]DCFPyL PET/CT scans can improve recurrence prediction based on clinical features alone. The highest model performance (C-index = 0.74 ± 0.13) was achieved by combining PET, CT, and clinical features. PET radiomics alone sometimes outperformed PET+Clinical models, suggesting that combining features may dilute biologically relevant signals due to feature redundancy (Table S2). Kaplan-Meier analysis confirmed the added prognostic value of radiomics signatures when stratifying patients into risk groups, particularly when multimodal features were used (Figure S1).

Model performance was highest in moderate-certainty negative scans (Figure 5), indicating that non-tumor radiomics captures subtle or systemic patterns of disease risk particularly in cases where visual interpretation is uncertain. The representative patients in Figure 6 illustrate this complementary role: a case predicted as high-risk despite a negative scan later developed nodal and osseous metastases, whereas a low-risk prediction remained recurrence-free on follow-up. Together, these findings suggest that non-tumor radiomics may provide clinically meaningful prognostic information precisely in visually ambiguous PSMA-negative cases, supporting its use alongside expert interpretation to guide surveillance and treatment decisions.

While our models effectively predicted overall progression risk, organ-specific radiomics did not accurately predict the site of future metastases. For example, models using only bone-derived features identified non-bone recurrence well but failed to detect patients who later developed bone metastases (mean ROC-AUC ~0.55), a pattern similarly observed for lung and pelvic tissues. This suggests that radiomics from common sites of recurrence reflect generalized disease susceptibility rather than site-specific risk.

Despite using nested CV (48) and ComBat harmonization, external validation showed moderate performance drops, especially for PET+CT and PET-only models, highlighting domain shift challenges. CT-only models were more stable, particularly at Center 3. These findings may reflect inherent differences between [$^{18}$F]DCFPyL (Center 1) and [$^{68}$Ga]PSMA-11 (Centers 2 and 3) tracers, such as resolution and biodistribution, which influence radiomics features and generalizability. Still, the modest drop (<10%) supports the feasibility of our framework, pending further multicenter validation. Consistent with patterns observed in other PET/CT radiomic models, external validation often reveals a performance drop exceeding 10 % (49, 50).

Limitations include potential selection bias, limited recurrence events, and absence of long-term ADT effects. Future studies should explore radiomics in patients with positive PSMA PET/CT, assess the role of pre-treatment tumor burden, and investigate how reconstruction parameters affect diagnostic certainty. Collectively, our results highlight the promise recurrence-prone radiomics in negative PSMA PET/CT scans as complementary biomarkers to improve risk stratification and guide personalized management in BCR patients.

## CONCLUSION

Radiomics features from common sites of recurrence on PSMA PET/CT-negative scans can predict progression in prostate cancer, offering prognostic value despite the absence of visible disease. Integrating clinical data and physician certainty improved model performance, with strongest predictive performance of non-tumor radiomics signature observed in moderate-certainty negative scans. While site-specific prediction was limited, global radiomics signatures reflected systemic risk. Prospective multicenter studies are needed to validate the idea of radiomics from common sites of recurrence for personalized management of PSMA-negative patients. Prospective trials are needed before using these models to guide treatment decisions such as surveillance versus salvage therapy.

# Supplementary Materials

## *Feature description*

**Table S1.** Feature description

| Feature Type | Features |
|---|---|
| *Shape* | Volume, Surface Area, Sphericity, Compactness, Maximum 3D Diameter |
| *Intensity Histogram (First-order features)* | Mean, Standard Deviation, Skewness, Kurtosis, Minimum, Maximum, 10th percentile, 90th percentile, Entropy |
| *GLCM (Gray Level Co-occurrence Matrix)* | Correlation, Contrast, Homogeneity, Energy, Entropy, Dissimilarity, Cluster Shade, Cluster Prominence, Difference Average, Difference Variance |
| *GLRLM (Gray Level Run Length Matrix)* | Short Run Emphasis (SRE), Long Run Emphasis (LRE), Gray-Level Non-Uniformity (GLNU), Run Length Non-Uniformity (RLNU), Run Percentage (RP), Low Gray-Level Run Emphasis (LGRE), High Gray-Level Run Emphasis (HGRE) |
| *GLSZM (Gray Level Size Zone Matrix)* | Small Zone Emphasis (SZE), Large Zone Emphasis (LZE), Gray-Level Non-Uniformity (GLNU), Zone Size Non-Uniformity (ZSNU), Zone Size Percentage (ZSP), Low Gray-Level Zone Emphasis (LGZE), High Gray-Level Zone Emphasis (HGZE) |
| *GLDM (Gray Level Dependence Matrix)* | Small Dependence Emphasis (SDE), Large Dependence Emphasis (LDE), Dependence Non-Uniformity (DNU), Gray-Level Non-Uniformity (GLNU), Dependence Entropy |
| *NGTDM (Neighborhood Gray Tone Difference Matrix)* | Coarseness, Contrast, Busyness, Complexity, Strength |
| *CSH (Cumulative SUV Histogram)* | Area Under Curve (AUC) |

Table S2. The results of using clinical features, non-tumor radiomics from PET and CT images and the combination of them for recurrence and time to recurrence prediction. Pre-PET ADT: adjuvant androgen deprivation therapy (ADT) prior to PET

| Feature set | Selected features | C-index | P <0.05 (Cox) |
|---|---|---|---|
| Clinical | Baseline PSA<br>T Category<br>Gleason<br>ISUP Grade Group<br>Velocity<br>Treatment Category<br>pre-PET ADT | 0.65±0.05 | Pre-PET ADT |
| PET | Lung GLCM Correlation<br>Lung INTENSITY Kurtosis<br>Lung GLCM Joint-Maximum<br>Vertebrae C2 GLSZM Normalised Zone Size Non-Uniformity<br>Vertebrae C4 INTENSITY HISTOGRAM Discretized intensity range<br>Vertebrae C4 GLSZM Small Zone Emphasis<br>Vertebrae L5 GLSZM Normalised Zone Size Non-Uniformity<br>Vertebrae T1 INTENSITY HISTOGRAM Discretized skewness<br>Vertebrae T5 NGTDM Busyness<br>Vertebrae T8 GLCM Difference Average | 0.72±0.10 | Lung INTENSITY Kurtosis<br>Vertebrae C4 GLSZM Small Zone High Grey Level Emphasis<br>Vertebrae T8 GLCM Difference Average |
| PET+Clinical | Clinical<br>Lung INTENSITY Kurtosis<br>Vertebrae C5 GLSZM Zone Size Entropy<br>Vertebrae L4 NGTDM Complexity<br>Hip GLSZM Small Zone Emphasis<br>Vertebrae C2 GLSZM Zone Size Non-Uniformity | 0.71±0.15 | Pre-PET ADT<br>Lung INTENSITY Kurtosis<br>Vertebrae L4 NGTDM Complexity |
| CT | Hip INTENSITY HISTOGRAM Minimum Histogram Gradient Grey Level<br>Hip GLSZM Grey Level Non-Uniformity<br>Lung GLSZM Zone Size Non-Uniformity<br>Lung NGTDM Complexity<br>Vertebrae C1 INTENSITY HISTOGRAM Minimum Histogram Gradient<br>Vertebrae C6 INTENSITY HISTOGRAM Intensity Histogram Coefficient of Variation<br>Vertebrae L4 INTENSITY HISTOGRAM Intensity Histogram Mean<br>Vertebrae L5 INTENSITY HISTOGRAM Maximum Histogram Gradient<br>Vertebrae T3 INTENSITY HISTOGRAM Maximum Histogram Gradient<br>Vertebrae T7 INTENSITY HISTOGRAM Maximum Histogram Gradient | 0.66±0.17 | Vertebrae C1 INTENSITY HISTOGRAM Minimum Histogram Gradient<br>Vertebrae L4 INTENSITY HISTOGRAM Minimum Histogram Gradient Grey Level |
| CT+Clinical | Clinical<br>Lung INTENSITY_HISTOGRAM 10th Percentile<br>Vertebrae C1 INTENSITY HISTOGRAM Minimum Histogram Gradient<br>Vertebrae L4 INTENSITY HISTOGRAM Minimum Histogram Gradient Grey Level<br>Vertebrae L5 INTENSITY HISTOGRAM Maximum Histogram Gradient<br>Hip left INTENSITY HISTOGRAM Intensity Histogram Mean | 0.71±0.14 | Pre-PET ADT<br>vertebrae C1 INTENSITY HISTOGRAM Minimum Histogram Gradient |
| PET+CT | Lung GLCM Correlation<br>Lung INTENSITY Kurtosis<br>Vertebrae C4 INTENSITY HISTOGRAM Range<br>Vertebrae L5 GLSZM Normalised Zone Size Non-Uniformity<br>Vertebrae T1 INTENSITY HISTOGRAM Quartile Coefficient of Dispersion<br>Vertebrae T1 GLCM Cluster Shade<br>Vertebrae T8 GLCM Difference Average<br>Vertebrae T9 INTENSITY HISTOGRAM Skewness<br>Vertebrae L4 INTENSITY HISTOGRAM Maximum Histogram Gradient Grey Level<br>Vertebrae T8 GLCM ClusterShade | 0.69±0.12 | Lung INTENSITY Kurtosis |
| PET+CT+Clinical | Clinical<br>Lung INTENSITY Kurtosis<br>Vertebrae C4 GLSZM Small Zone Emphasis<br>Vertebrae L4 NGTDM Complexity<br>Prostate/bed NGTDM Complexity<br>Vertebrae L4 GLCM ClusterShade | 0.74±0.13 | Pre-PET ADT<br>Lung INTENSITY Kurtosis<br>Vertebrae L4 NGTDM Complexity |

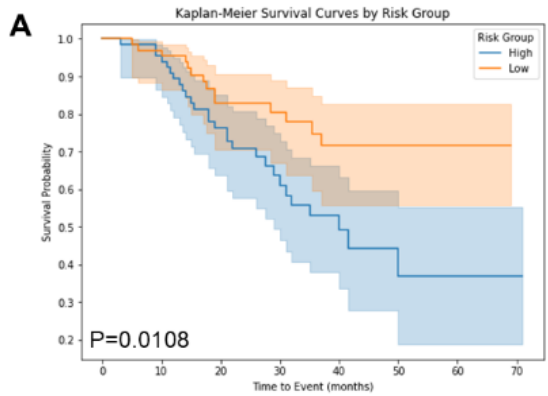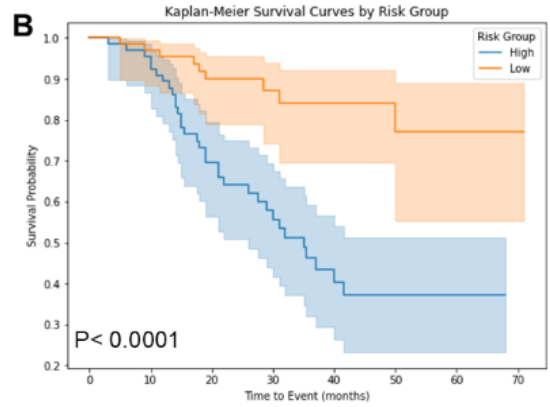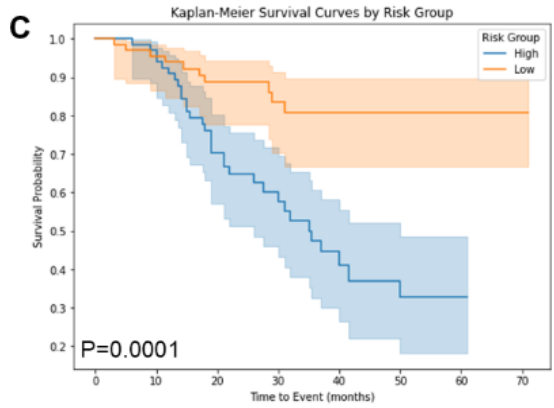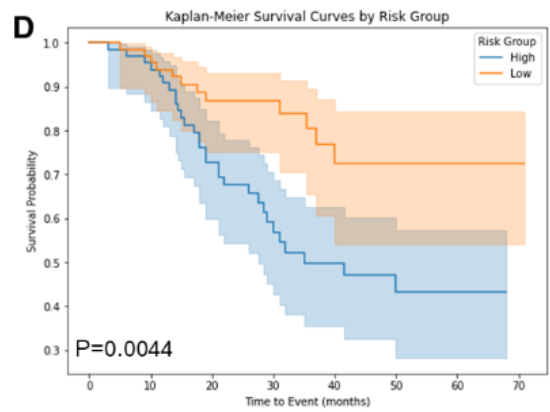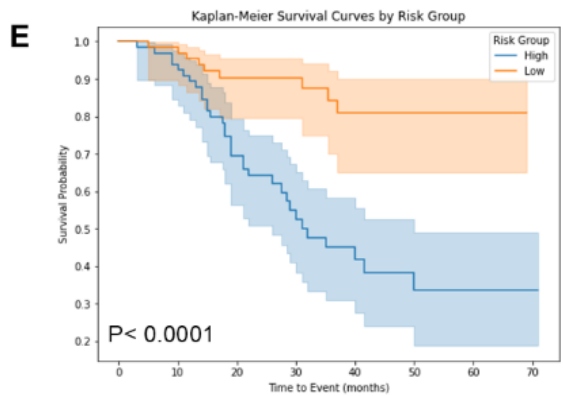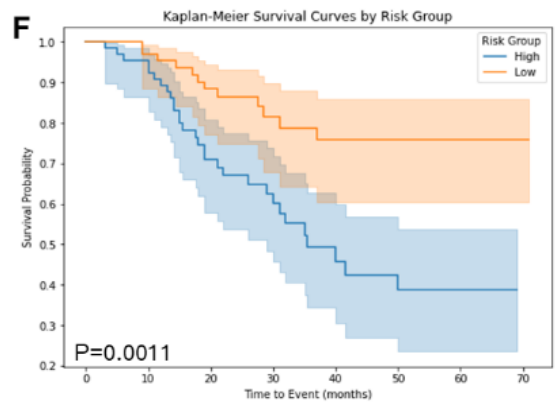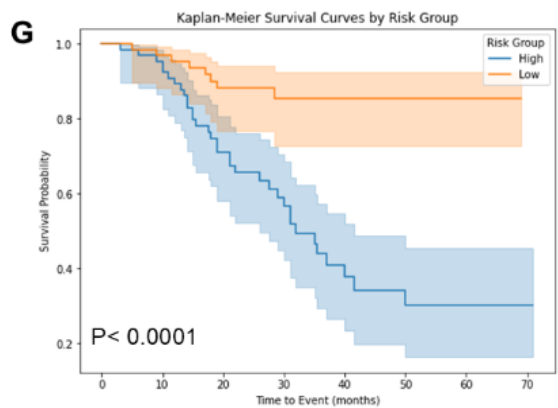

Figure S1. Kaplan–Meier survival curves for disease-free survival according to high- and low-risk groups derived from different feature sets. Risk groups were defined by the median prognostic score from the respective multivariable Cox proportional hazards models. (A) Clinical features, (B) PET features, (C) PET+Clinical features, (D) CT features, (E) CT+Clinical features, (F) PET+CT features, and (G) PET+CT+Clinical features. Shaded areas represent the 95% confidence intervals for survival probability. Corresponding log-rank p-values are shown on each panel.

*External validation*

The PET-negative cases were selected among patients from two external centers (2: IKH and 3: RH) who had experienced biochemical recurrence following curative-intent therapy, including RP, RT, or both, with or without ADT, and who received no additional treatment after their negative scan.

**Center 2:** [$^{68}$Ga]PSMA PET/CT scans (n=6) were acquired on a GE Discovery IQ system (matrix: 192×192; voxel size: 3.26 mm³; VPHDS reconstruction (GE hybrid reconstruction approach); model-based scatter correction), with weight-adjusted doses of 1.85–2.54 MBq/kg (total dose: 130–241 MBq).

**Center 3:** [$^{68}$Ga]PSMA PET/CT scans (n=21) were acquired on a Siemens Biograph mCT system (matrix: 168×168; voxel size: 5 mm³; Point spread function (PSF) reconstruction; model-based scatter correction), with weight-adjusted doses of 0.02–2.79 MBq/kg (total dose: 1.42–178.71 MBq).

Follow-up time was 15.2 months (range: 11.3–33.1 months) for Center 2 and 15.6 months (range: 0.9–35.8 months) for Center 3.

*Organ segmentation*

TotalSegmentator [1], built on the nnU-Net framework [2], is an auto-configuring deep learning tool for 3D CT segmentation that eliminates manual tuning. It dynamically adjusts parameters for diverse imaging scenarios. Trained on 104 annotated structures using extensive data augmentation, it robustly generalizes [1]. Its self-configuring pipeline automatically sets patch size, batch size, and network architecture. Validated on unseen CT scans, the model accurately segments regions such as the hip bone, liver, lung, sacrum, vertebrae, prostate, and prostate bed, demonstrating reliable performance for a wide range of clinical applications [1] even in low dose CT images[3].

### *PET/CT registration*

Although PET/CT scans are inherently co-registered, we applied additional image registration using Elastix [4] to enhance the alignment between the PET and CT images for each patient based on deformable registration that employs a B-spline transform with the mutual information similarity metric. This process allowed for the precise registration of the segmented organs from the CT scans to the corresponding PET scans.